\documentstyle[prl,aps,amsmath,psfig]{revtex}
\begin{document}
\draft
\twocolumn[\hsize\textwidth\columnwidth\hsize\csname
@twocolumnfalse\endcsname

\widetext

\title{Quantum Phase Transition in Coupled Spin Ladders}
\author{Luca Capriotti$^{1}$\cite{lc} and Federico Becca$^{2}$} 
\address{
    ${^1}$ Istituto Nazionale per la Fisica della Materia, 
    Unit\`a di Ricerca di Firenze, I-50125 Firenze, Italy \\
    ${^2}$ Institut de Physique Th\'eorique, Universit\'e de Lausanne, CH-1015 Lausanne, Switzerland} 
    
\date{\today}
\maketitle
\begin{abstract}
The ground state of an array of coupled, spin-half, antiferromagnetic ladders 
is studied using spin-wave theory, exact diagonalization (up to 36 sites) and
quantum Monte Carlo techniques (up to 256 sites).
Our results clearly indicate the occurrence of a zero-temperature phase transition
between a N\'eel ordered and a non-magnetic phase at a finite value of the inter-ladder coupling
($\alpha_c\simeq0.3$). This transition is marked by remarkable changes in the structure of the excitation 
spectrum. 
\end{abstract}

]
\narrowtext


Ground-state (GS) correlations in a quantum antiferromagnet are 
closely related to the nature of low-energy excitations.
This is deeply connected to the mechanism of spontaneously symmetry breaking and turns 
out clearly whenever 
competing GS's give rise to a quantum
phase transition, leading in general to remarkable changes of the excitation spectrum 
at the critical point \cite{sachdev}. 

\begin{figure}
\centerline{\psfig{bbllx=225pt,bblly=210pt,bburx=425pt,bbury=402pt,%
figure=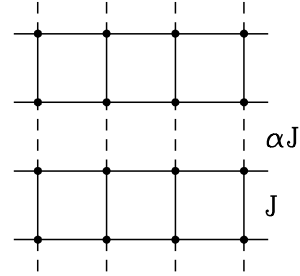,width=40mm,angle=0}}
\caption{\baselineskip .185in \label{latt}
The array of coupled two-leg ladders. 
}
\end{figure}

A simple model system experiencing such changes in the structure of low-energy 
excitations in correspondence of a quantum phase transition
is the spin-half Heisenberg antiferromagnet
on a two-dimensional array of coupled ladders:
\begin{equation} 
\hat{\cal{H}}=J\sum_{r}
\hat{{\bf {S}}}_{r} \cdot \hat{{\bf {S}}}_{r+\hat{x}}
+ \sum_{r}J_r
\hat{{\bf {S}}}_{r} \cdot \hat{{\bf {S}}}_{r+\hat{y}}~,
\label{hamilt}
\end{equation}
where ${\bf \hat{S}}_{r}=(\hat{S}^x_r,\hat{S}^y_r,\hat{S}^z_r)$
are $s=1/2$ operators on the sites $r=(r_x,r_y)$ of a $L\times L$ lattice
with periodic boundary conditions; $\hat{x}=(1,0)$, $\hat{y}=(0,1)$, 
$J_r=J$ or $J_r=\alpha J$ ($J>0$), depending on the parity of $r_y$, and $\alpha$ is
the inter-ladder coupling (see Fig.~\ref{latt}). 
Such Hamiltonian interpolates between the Heisenberg model
on the square lattice ($\alpha=1$) and a system of $L/2$ decoupled two-leg ladders 
($\alpha=0$). 
In the square lattice limit, the GS has N\'eel long-range order, 
with a gapless excitation spectrum and a sizable value of the antiferromagnetic order 
parameter \cite{heisenberg,matteuccio}. The two-leg ladder, instead, has a finite triplet 
gap in the thermodynamic limit and no long-range antiferromagnetic 
order \cite{twoleg}. 
As a result a quantum critical point between a gapless,
magnetically ordered phase and a non-magnetic GS of purely quantum mechanical nature 
is expected at a critical value of the inter-ladder coupling, $\alpha_c$ \cite{sachdev}. 

Besides its intrinsic theoretical interest, this model has been recently studied
due to the discovery of several compounds, such as ${\rm SrCu}_2{\rm O}_3$ and 
$({\rm VO})_{2}{\rm P}_{2}{\rm O}_7$, displaying clear signatures of a finite gap ($\sim J/2$)
in the excitation spectrum related to the underlying ladder structure \cite{imada,azzouz}.
In addition, the coupled-ladder Hamiltonian (\ref{hamilt}) has been also considered as a 
simplified model for the magnetism of the striped phase in ${\rm CuO}_2$ planes 
of hole-doped $high{-}T_c$  copper-oxides \cite{stripes1,stripes2}.

In this paper, by means of spin-wave (SW) theory, exact diagonalization by the Lanczos method 
and zero-temperature quantum Monte Carlo techniques, including variational and Green function Monte Carlo 
(GFMC) \cite{trivedi,matteuccio}, we focus on the GS and the low-energy excitations of the 
coupled spin-ladder model (\ref{hamilt}). 
In particular, by means of a systematic size-scaling 
analysis, we will show how the structure of low-energy excitation spectrum
provides  very clear indications of the changes of the GS state correlations occurring at
a critical point, thus allowing us to put on firmer grounds the existence of a quantum phase 
transition at $\alpha_c \simeq 0.3$.

\begin{figure}
\centerline{\psfig{bbllx=45pt,bblly=405pt,bburx=525pt,bbury=680pt,%
figure=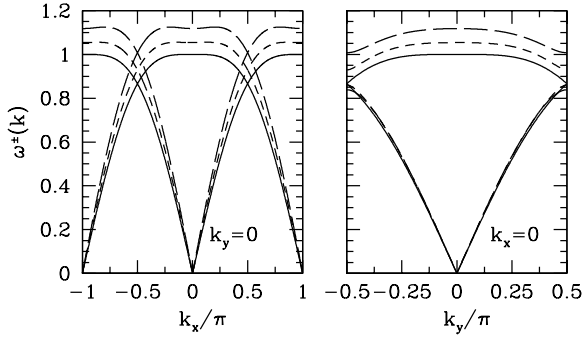,width=80mm,angle=0}}
\caption{\baselineskip .185in \label{omega}
Linear SW dispersion relation for $\alpha=1.0$ (continuous line),
$\alpha=0.8$ (short dashes), and $\alpha=0.6$ (long dashes).
}
\end{figure}

The simplest approach to the study the effects of zero-point fluctuations on 
the GS correlations of a quantum magnet 
is SW theory. This has turned out to be a very reliable
approximation of the GS of spin-half systems whenever it has 
long-range antiferromagnetic order even in presence of strong quantum fluctuations 
\cite{swtriang}.
By the use of the standard Holstein-Primakoff representation of the spin operators
we can compute the fluctuations over the classical solution at the leading order in $1/s$.
In contrast to the Heisenberg model on the square lattice, the reduced translation 
symmetry along the $y$ direction, implies the existence of two inequivalent branches of SW excitations,
and the energy of such SW modes, $\omega_{k}^\pm$, reads
\begin{equation}
\omega_{k}^\pm=[2D^2-(\gamma_{k}^2+\gamma_{{\bar {k}}}^2)-2\delta_{k}^2\pm F_{k}]^{1/2}/\sqrt{2}~,
\end{equation}
with $F_{k}=[(\gamma_{k}^2-\gamma_{{\bar {k}}}^2)^2+4\delta_{k}^2
(\gamma_{k}+\gamma_{{\bar {k}}})^2]^{1/2}$, $\gamma_{k} = (\beta \cos k_x + \cos k_y)/2$,
$\delta_{k}=(1-\alpha)/2(1+\alpha)\sin k_y$,
${\bar {k}} = {k} + (0,\pi)$, $D=(1+\beta)/2$, and $\beta=2/(1+\alpha)$. 
The dispersion relation of the two SW modes is plotted in Fig.~\ref{omega}. 
Notice that for $\alpha <1$ a finite gap develops between the optical and the acoustical branch of the
SW dispersion along the $y$-direction due to the reduced translation symmetry.
The acoustical branch remains always gapless at $k=(0,0)$ and $k=(\pi,0)$, 
corresponding to the Goldstone modes
(in a reduced-zone scheme) associated with the SU(2) symmetry-breaking assumption.

\begin{figure}
\centerline{\psfig{bbllx=50pt,bblly=205pt,bburx=570pt,bbury=630pt,%
figure=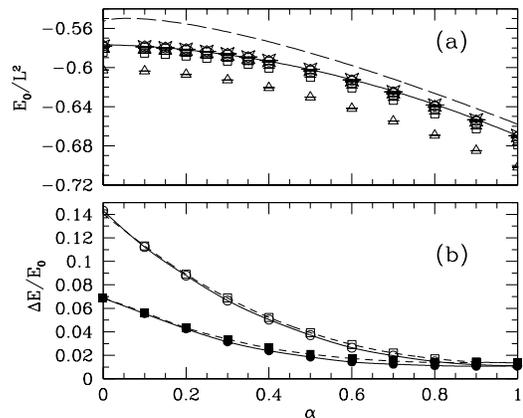,width=70mm,angle=0}}
\caption{\baselineskip .185in \label{accuracy}
(a): $\alpha$-dependence of the exact GS energy per site 
for $L=4$ (triangles), 6 (squares), 10 (pentagons), and 12 (circles).
Stars and continuous line are bulk-limit extrapolations
and the dashed line is the linear SW prediction.
(b): accuracy of the variational energy of the SW-like
wave function of Eq.~(\protect\ref{eq.gwfj1j2}) with the isotropic
Jastrow potential of Ref.~\protect\cite{franjo} (empty symbols) and
with its generalization to coupled ladders (full symbols).
$6\times 6$: squares; $12\times 12$: circles.
}
\end{figure}

The expansion of the staggered magnetization at the first order in $1/s$ indicates 
the stability of the N\'eel order up to very small values of the inter-ladder coupling  
($\alpha_c\simeq 0.01$, for $s=1/2$). Within the same approximation the GS energy per site 
reads
\begin{equation}
{E_0^{SW} \over L^2}=-\tilde{J} s(s+1) (1+\beta) +  {2\tilde{J} s  \over  L^2} \sum_{k}^{RBZ} 
(\omega_{k}^++\omega_{k}^-)~,
\end{equation}
where $\tilde{J} = (1+\alpha)J /2$ and the summation is over the $k$-vectors belonging to
the sector of the Brillouin zone with $-\pi/2<k_y\leq \pi/2$.
The SW energy in the thermodynamic limit as a function of the
inter-ladder coupling $\alpha$ is plotted in Fig.~\ref{accuracy}-a.
For spin models the exact value of the GS energy can be calculated
numerically for small clusters with the Lanczos technique and for rather larger sizes, in absence of frustration,
with the GFMC method \cite{trivedi,matteuccio}.
The comparison with the extrapolation to the bulk limit of the GFMC results
(up to $L\leq 16$, see below for details) 
indicates that the latter analytical approach provides 
accurate estimates of the GS energy only in the regime of large 
inter-ladder coupling ($\alpha\gtrsim 0.6$) thus suggesting that the SW approach
underestimates the actual effect of quantum fluctuations and, therefore, the critical value of
the inter-ladder coupling.

By means of the SW analysis it is also possible to derive a variational wave function 
providing a good representation of the GS and low-lying excited 
states at least for $\alpha\to 1$, when long-range antiferromagnetic correlations are expected. 
This wave function, which is also easily computable when used for importance sampling 
in a GFMC calculation to reduce the numerical effort \cite{trivedi,matteuccio}, 
can be obtained starting from a N\'eel ordered state and including Gaussian fluctuations 
by means of a Jastrow factor \cite{franjo}:
\begin{equation}
\label{eq.gwfj1j2}
|AF\rangle={\cal P}_{S}\sum_{x} S_M(x) \exp{\Big[ \frac{1}{2} 
\sum_{r,r^\prime} v(r-r^\prime)S_{r}^zS_{r^\prime}^z\Big]}|x\rangle~.
\end{equation}
Here $|x\rangle$ is an Ising spin configuration specified by assigning
the value of $S_r^z$ for each site, ${\cal P}_S$ is the projector onto
the subspace with $S^z_{tot}=\sum_{r}S^z_r=S$, and $S_M(x)=(-1)^{N_\uparrow(x)}$ is
the {\em Marshall sign}, reproducing exactly the phases of the GS \cite{lieb}.  
This depends only on the number of up spins on one of the two sublattices, $N_\uparrow(x)$, 
so that $|N\rangle=\sum_x S_M(x)|x\rangle$ represents the classical N\'eel state \cite{swtriang}.  
For the two-body Jastrow potential, the simple form,
based on the consistency with linear SW theory in the square lattice case\cite{franjo},  
$v(r)= (\eta/L^2)\sum_{k\ne0} e^{-ik\cdot r} v_k$, with 
$v_k=1-\sqrt{(1+\Gamma_k)/(1-\Gamma_k)}$,  $\Gamma_k=(\cos k_x + \cos k_y)/2$, and $\eta$ 
variational parameter,  provides good variational estimates 
only for values of $\alpha$ very close to 1 (see Fig.~\ref{accuracy}-b),
as it is expected since the  modulation of the exchange interaction weakens 
the antiferromagnetic ordering and enhances spin fluctuations.

Fairly better variational results can be obtained by generalizing the 
latter wave function to the case of coupled spin ladders using 
the results of the SW analysis presented above. 
Indeed, the resulting wave function, whose explicit expression is too cumbersome 
to be reported here, provides very good variational 
estimates up to remarkably smaller values of $\alpha\sim 0.6$ (Fig.~\ref{accuracy}-b). Decreasing further the 
inter-ladder coupling the accuracy of the SW variational wave function rapidly decreases. 
This clearly indicates that entering the small-$\alpha$ regime 
our gapless N\'eel ordered variational {\em ansatz} does not reproduce correctly 
the GS correlations.

\begin{figure}
\centerline{\psfig{bbllx=35pt,bblly=180pt,bburx=555pt,bbury=610pt,%
figure=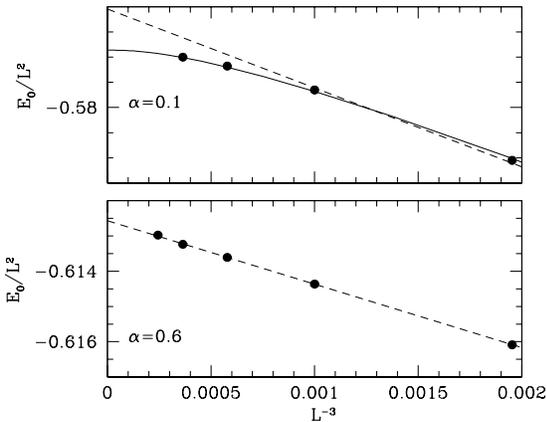,width=70mm,angle=0}}
\caption{\baselineskip .185in \label{erg}
Size scaling of the GS energy per site  for $\alpha=0.1$ and $\alpha=0.6$.
The dashed line is the linear fit of the data for $L\leq 10$
and the continuous line is the fit according to Eq.~(\protect\ref{experg}).
}
\end{figure}   

Useful indications on the nature of the thermodynamic GS
can be obtained numerically by studying the finite-size scaling of the GS energy,
which is deeply connected to the nature of the excitation spectrum and therefore
to possible thermodynamic broken symmetries \cite{bernu}.
In fact, in presence of N\'eel long-range order, being the spectrum gapless and the magnon
dispersion relation linear in the wave vector ${k}$,
the leading finite-size correction to the GS energy
per site, $e_0(L)=E_0(L)/L^2$, is ${\cal O}(L^{-3})$\cite{ziman}. 
Instead, in presence of a finite correlation length, a finite gap in the spin 
excitation spectrum and an exponential asymptotic dependence is expected. 
Here we have assumed a size dependence of the form 
\begin{equation}
e_0(L)=e_0(\infty)+c_0\exp(-L/L_0)/L^2~,
\label{experg}
\end{equation}
which has been employed in Ref.~\cite{dagotto} for the two-leg ladder.
As it is shown in Fig.~\ref{erg}, the size-scaling law predicted for a long-range 
ordered GS is fulfilled for values of $\alpha$ close enough to 1 
while for small value of $\alpha$ a clear deviation from this behavior
is observed and the exponential law (\ref{experg}) is instead satisfied.
This provides another numerical evidence of the melting of the antiferromagnetic 
long-range order due to the transition to the ladder-like regime.

\begin{figure}
\centerline{\psfig{bbllx=15pt,bblly=160pt,bburx=530pt,bbury=610pt,%
figure=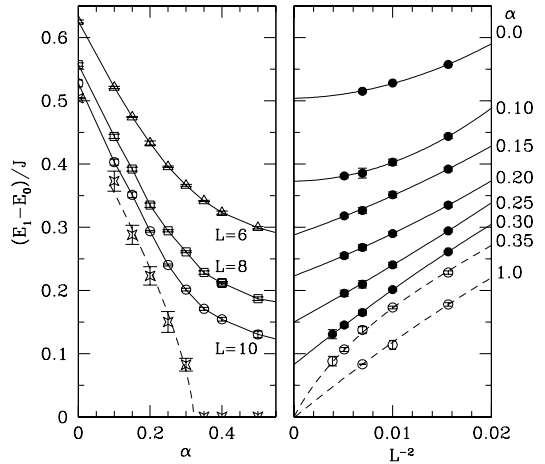,width=70mm,angle=0}}
\caption{\baselineskip .185in \label{gap}
Left panel: $\alpha$-dependence of the triplet gap for various lattice sizes.
Stars are bulk-limit extrapolations according to
Eq.~(\protect\ref{gapext}) and the dashed line is a fit according to $\Delta \propto (\alpha_c-\alpha)^{0.69}$.
Right panel: size scaling of the triplet gap for different values of $\alpha$.
Continuous  and dashed lines are fits according to Eqs.~(\protect\ref{gapext}) and (\protect\ref{gapheis}),
respectively.
}
\end{figure}

The occurrence of a quantum phase transition to a non-magnetic GS is
clearly confirmed by the study of the spin gap shown in Fig.~\ref{gap}.
This physical quantity can be measured straightforwardly with GFMC 
by performing two different simulations in the $S^z_{tot}=0$ and $S^z_{tot}=1$ 
subspaces. In contrast to finite-temperature algorithms, the calculation
of the spin gap within GFMC does not involve any fitting procedure 
of low-temperature data and it is exact within statistical error.
As shown in the left panel of Fig.~\ref{gap}, the triplet gap, 
$\Delta=E_1-E_0$, is a 
decreasing function of the inter-ladder coupling even if,  on finite-sizes,
this quantity is always non-zero for any value of $\alpha$. 
However, the size scaling of the spin gap, shown
in the right panel of the same figure, indicates a clear deviation
from the size dependence expected in presence of long-range N\'eel order \cite{ziman},
\begin{equation}
\Delta_L=a/L^2+b/L^3~,
\label{gapheis}
\end{equation}
and the opening of a finite gap in the thermodynamic excitation spectrum 
for $\alpha \lesssim 0.35$. In this regime the extrapolation to 
the bulk limit of the finite-size data can be done using a law
of the type\cite{white}:  
\begin{equation}
\Delta_L=\Delta+a/L^2+b/L^4~.
\label{gapext}
\end{equation}

An estimate of the critical value of the inter-ladder coupling
can be obtained by fitting the extrapolated values of the gap
with the scaling law $\Delta=(\alpha_c-\alpha)^\nu$, with $\nu \simeq 0.69$, 
predicted for a quantum phase transition in (2+1) dimensions\cite{sachdev}.
This procedure gives the value $\alpha_c\simeq 0.32\pm 0.03$, in agreement 
with previous numerical estimates obtained with
finite-temperature algorithms \cite{imada,stripes1} and the mean-field predictions 
of Ref.~\cite{gopalan} but in contrast with the conclusions of bond-mean-field theory \cite{azzouz} 
indicating the vanishing of the gap for infinitesimal values $\alpha$.
In contrast to frustrated systems like the $J_{1}{-}J_{2}$ model, 
where the SW theory provides an accurate prediction of the transition 
to a non-magnetic phase due to competing interactions \cite{swtriang,doucot,j1j2}, 
in this case the SW result
($\alpha_c \sim 0.01$) grossly underestimate the exact one. This can be ascribed to
the fact that in this model there are no competing GS's at the classical level, 
the N\'eel state being stable up to $\alpha=0$, and the transition has therefore 
a pure quantum origin. 
Within the $1/s$ expansion, instead, the SW velocity 
remains finite up to $\alpha=0$ and the reduction of the staggered magnetization 
is only due to the crossover to the one-dimensional regime in which 
antiferromagnetic long-range order is unstable.

\begin{figure}
\centerline{\psfig{bbllx=55pt,bblly=175pt,bburx=570pt,bbury=675pt,%
figure=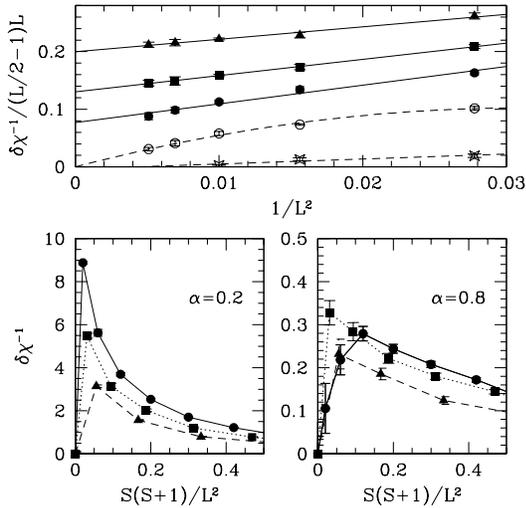,width=70mm,angle=0}}
\caption{\baselineskip .185in \label{tower}
Lower panels: rigid rotator anomaly vs $S$ for $\alpha=0.2$,
and $\alpha=0.8$: $L=6$ (triangles), $8$ (squares), $10$ (circles).
Upper panel: size scaling of $\delta\chi^{-1}(1)/(L/2-1)L$ for (from the top) $\alpha$=0.2, 0.25, 0.3, 0.4, and 0.8.
Lines are weighted quadratic fits.
}
\end{figure} 

A major fingerprint of the dramatic changes in the structure of the excitation
spectrum occurring in correspondence of the above quantum phase transition 
can be found in the finite-size behavior of the so called
{\em rigid rotator anomaly} \cite{bubbles}, $\delta \chi^{-1}(S) = 1/2\chi_S-1/2\chi_L$, with
\begin{equation}
\frac{1}{2\chi_S}= L^2\frac{E_S-E_0}{S(S+1)}~,
\end{equation}    
where $E_S$ is the energy of the lowest excitation with spin  $S$.
In fact, in presence of long-range antiferromagnetic order, 
the low-lying excited states of spin $S$ are predicted to behave as
the spectrum of a free quantum rotator, 
$ E_S-E_0 \propto S(S+1)/L^2$, as long as $S \ll L^2$\cite{bernu}. In contrast, 
the expected behavior for a spin ladder is
$E_S-E_0 \propto S$, as it is easy to understand in a spin-liquid Resonating Valence Bond (RVB) 
picture\cite{twoleg,white,rainbow}. 
As a result, in the gapless phase
the rigid rotator anomaly $\delta\chi^{-1}$ has to vanish identically in the thermodynamic limit
while in the gapped regime $\delta\chi^{-1}(S)\propto L(L/(S+1)-1)$ diverges linearly
with the volume. These features of the excitation spectrum clearly discriminate 
the two different zero-temperature phases of the present model, as it is evident from
Fig.~\ref{tower} displaying the very different behavior of $\delta\chi^{-1}$
below and above the critical point.

In summary, we have investigated the ground-state properties of an array
of coupled spin-half ladders using spin-wave theory and a numerical analysis
of the finite-size low-energy excitation spectrum.
While the spin-wave theory turns out to be reliable only in the regime of weak inter-ladder
coupling, our numerical results provide robust indications of the opening of a finite 
spin gap  in the thermodynamic limit for $\alpha \lesssim 0.3$,
corresponding to a quantum phase transition between a gapless N\'eel ordered phase and 
a spin-liquid RVB ground state.

We are deeply indebted to Alberto Parola and Sandro Sorella for many valuable 
discussions and for their constant encouragement.
Thanks also to Frederic Mila and Ingo Peschel for suggestions and useful discussions. 
This work was partially supported by MURST (COFIN00). 


\end{document}